# Generation of Cooperative Perception Messages for Connected and Automated Vehicles

Gokulnath Thandavarayan, Miguel Sepulcre, Javier Gozalvez

*Abstract*— Connected and Automated Vehicles (CAVs) utilize a variety of onboard sensors to sense their surrounding environment. CAVs can improve their perception capabilities if vehicles exchange information about what they sense using V2X communications. This is known as cooperative or collective perception (or sensing). A frequent transmission of collective perception messages could improve the perception capabilities of CAVs. However, this improvement can be compromised if vehicles generate too many messages and saturate the communications channel. An important aspect is then when vehicles should generate the perception messages. ETSI has proposed the first set of message generation rules for collective perception. These rules define when vehicles should generate collective perception messages and what should be their content. We show that the current rules generate a high number of collective perception messages with information about a small number of detected objects. This results in an inefficient use of the communication channel that reduces the effectiveness of collective perception. We address this challenge and propose an improved algorithm that modifies the generation of collective perception messages. We demonstrate that the proposed solution improves the reliability of V2X communication and the perception of CAVs.

## I. INTRODUCTION

Automated vehicles utilize onboard sensors to perceive the surrounding environment and drive autonomously. The perception capabilities of these sensors can be limited for example due to the presence of obstacles (including other vehicles) or adverse weather conditions. Connected and Automated Vehicles (CAVs) can improve their perception capabilities if vehicles exchange information about what they sense using V2X communications. Vehicles can use the exchanged information to detect vehicles or objects that were not detected by their onboard sensors. This is known as cooperative or collective perception. Previous studies [1] have identified the potential of cooperative perception to improve the vehicles' perception beyond the sensors' detection range.

First collective or cooperative perception studies analyzed the advantages and disadvantages of exchanging raw sensor data, processed metadata or compressed data [2]. Exchanging raw sensor data would require large communication bandwidths that cannot be guaranteed by existing technologies (such as DSRC, ITS-G5 or C-V2X) when the network scales. Recent studies have focused on the exchange of information about detected objects including their position, speed and size. For example, the study in [3] compares the perception achieved when the information about the detected objects is attached to existing Cooperative Awareness Messages (CAMs) or is transmitted in separate messages.

Other studies seek to control the information exchanged between vehicles in order to reduce the load on the communications channel. In [4], authors propose that each vehicle should transmit the information about a detected object only if this information is valuable for its neighboring vehicles. Accurately estimating the value of the information in a distributed and highly dynamic environment is a significant challenge. In [5], the same authors partially address this challenge by using deep reinforcement learning to select the information to be transmitted. In [6], authors propose a method to reduce the channel load by transmitting only the most relevant information. This method takes into account the area covered by the sensors that is not covered by nearby vehicles. The work in [7] proposes an analytical performance model for collective perception. The study in [8] shows that existing rules to generate collective perception messages can generate a lot of redundant information in the network as vehicles receive many updates per second about a detected object. Authors propose in [8] a method to reduce this redundancy in order to improve the networks' scalability. Additional redundancy mitigation mechanisms were proposed in [9]. The study in [10] analyzes different content control schemes that decide whether to report or not about certain detected objects based on their distance to the sender vehicle and their impact on position tracking errors. The study determines that objects that are located farther away from the sender but near the edge of the sensors' range should be prioritized. These studies show the need to control the exchanged information without degrading the perception.

The perception also depends on how frequently collective perception messages are generated and transmitted. In principle, a frequent transmission of collective perception messages could improve the perception of CAVs. However, this can be compromised if vehicles generate too many

This work has been partly funded by the European Commission through the TransAID project under H2020 Programme, Grant Agreement no. 723390.

Gokulnath Thandavarayan, Miguel Sepulcre and Javier Gozalvez are with Universidad Miguel Hernandez de Elche (UMH), Spain. (e-mail: gthandavarayan@umh.es, msepulcre@umh.es, j.gozalvez@umh.es).

messages and saturate the communications channel. An important aspect is then when vehicles should generate the perception messages. ETSI (European Telecommunications Standards Institute) has proposed to date the first set of message generation rules for collective perception [11]. These rules define when vehicles should generate collective perception messages and what should be their content. [12] showed that ETSI generation rules result in the frequent transmission of collective perception messages with information about a small number of detected objects. This results in an inefficient use of the communication channel due to the frequent transmission of packet headers. Overloading the communication channel with frequent messages can also decrease the packet delivery ratio and therefore the effectiveness of cooperative perception. This paper addresses these challenges with an improved algorithm that modifies the generation of collective perception messages and reorganizes their content. To our knowledge, this is the first study that tackles the problem of generating frequent collective perception messages reporting about a small number of objects. The proposal is referred to as *look-ahead* and an earlier version was included in [11]. It modifies the ETSI generation rules to reorganize the transmission of objects in collective perception messages. The reorganization results in that vehicles transmit less messages, and each message includes information about a higher number of detected objects. The proposed solution reduces the channel load and improves the reliability of V2X communications and the perception capabilities of CAVs.

## II. COLLECTIVE PERCEPTION SERVICE

ETSI has recently approved a Technical Report that defines the so-called Collective Perception Service (CPS) [11]. The report presents the first proposal to standardize the Collective Perception Message (CPM) format and the CPM generation rules[1]. A CPM contains information about the vehicle that generates the CPM, its onboard sensors (their range, field of view, etc.), and the detected objects (position, speed, size, etc.). In particular, CPM messages include an ITS (Intelligent Transport Systems) PDU (Protocol Data Unit) header and 5 containers: Management Container, Station Data Container, Sensor Information Containers (SICs), Perceived Object Containers (POCs) and Free Space Addendum Container (FSAC). The ITS PDU header includes data elements such as protocol version, the message ID and the Station ID. The Management Container is mandatory and provides basic information about the transmitter, including its type (e.g. vehicle or RSU) and position. The Station Data Container is optional and includes additional information about the transmitter (e.g. its speed, heading, or acceleration). The SIC is optional and can report up to 128 sensors in a CPM. These containers describe the capabilities of the sensors embedded in the transmitting vehicle. The POCs is optional and can report up to 128 detected objects in a CPM. A POC provides information about the detected objects (e.g. their distance to the transmitting vehicle, speed and dimensions), and the time at which the measurements were done. The FSAC is optional and describes the free space areas within the sensor detection areas.

The CPM generation rules define how often a vehicle should generate and transmit a CPM and the information it should include. The current ETSI CPM generation rules [11] establish that a vehicle has to check every $T\_GenCpm$ if a new CPM should be generated and transmitted, with $0.1s \leq T\_GenCpm \leq 1s$. A vehicle should generate a new CPM if it has detected a new object, or if any of the following conditions are satisfied for any of the previously detected objects:

1. Its absolute position has changed by more than 4m since the last time its information was included in a CPM.
2. Its absolute speed has changed by more than 0.5m/s since the last time its information was included in a CPM.
3. The last time the detected object was included in a CPM was 1 (or more) seconds ago.

A vehicle includes in a new CPM all new detected objects and those objects that satisfy at least one of the previous conditions. The vehicle still generates a CPM every second even if none of the detected objects satisfy any of the previous conditions. The information about the onboard sensors is included in the CPM only once per second.

ETSI has proposed to date the first set of generation rules for collective perception. These rules are then considered as benchmark and we next analyze their performance to identify existing challenges and motivate our proposal.

## III. PROBLEM STATEMENT

Let's consider the scenario in Figure 1 where an ego vehicle has 6 neighboring vehicles. Let's assume that the ego vehicle is equipped with a sensor that has a Field of View (FoV) of 360º and all vehicles move at 70 km/h. The ego vehicle generates CPMs following the current ETSI CPM generation rules and checks the conditions to generate a CPM every $T\_GenCpm$=0.1 s. As a result, the ego vehicle includes each detected vehicle in a CPM every 300 ms. Let's suppose, as an example, a scenario where the ego vehicle detects for the first time all neighboring vehicles in a time interval $\tau \leq 0.1$ s. In this scenario (Scenario 1), the ego vehicle generates one CPM every 300 ms, and each CPM includes the information of the 6 detected vehicles (see Scenario 1 in Figure 1). It is though very unlikely that an ego vehicle can detect all its neighboring vehicles in the same time interval. In a more realistic scenario, vehicles constantly enter and leave the sensor detection range of an ego vehicle at different times. The ego vehicle will then include the detected objects (i.e. vehicles) in different CPMs. Let's consider in Scenario 2 that the ego vehicle detects two different neighboring vehicles in every time interval $\tau = 0.1$ s. In this scenario, the ego vehicle ends up transmitting one CPM every 100 ms instead of every 300 ms like in Scenario 1. In Scenario 2, each CPM includes now information about 2 detected objects every 100 ms instead of 6 every 300 ms (see Scenario 2 in Figure 1). Transmitting more CPMs per second consumes more bandwidth since each CPM includes the ITS PDU Header, the Management and Station Data containers. They occupy around 121 Bytes and are shown in grey color in Figure 1. In addition, each CPM generates protocol headers from the Transport, Network, MAC (Medium Access Control) and PHY (Physical) layers. They occupy around 80 Bytes and are shown in blue color in Figure 1. Figure 1 clearly shows that the transmission of more CPMs with information about less objects (Scenario 2) increases the signaling overhead compared to transmitting less CPMs that contain a larger number of objects (Scenario 1).

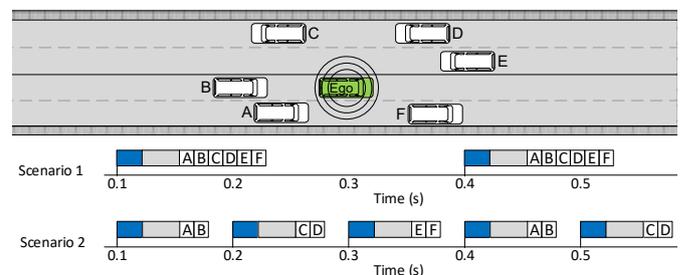

Figure 1. Example to illustrate the problem statement.

---
[1] The Technical Report in [11] will serve as a baseline for the specification of CPS in ETSI TS 103 324, which has not yet been approved, so the current CPM message format and generation rules are still a proposal.

We have analyzed and quantified the effects illustrated in the example in Figure 1 by means of simulating an urban and a highway scenario. These simulations consider realistic conditions where the sensors embedded in the vehicles detect the objects and the CPMs are generated following the conditions defined in Section II. For the highway scenario, simulations have been conducted for a 5 km long six-lane highway. We simulated two traffic densities following [13]: 120 veh/km (high density) and 60 veh/km (low density). We configured different speeds per lane to statistically mimic a typical 3-lane US highway. The speed of lanes varies between 118 km/h and 140 km/h for the low traffic density scenario and between 59 km/h and 70 km/h for the high traffic density scenario. For the urban scenario, a Manhattan-like grid scenario with 9x7 blocks is simulated. The size of each block is 433 m x 250 m and each street has 4 lanes [13]. In this scenario, the maximum speed is 70 km/h and two traffic densities are considered: 25 veh/km (low density) and 45 veh/km (high density). In both urban and highway scenarios, the mobility of vehicles is simulated with the road mobility simulator SUMO. To avoid boundary effects, statistics are only collected in the 2 km road segment around the middle of the highway scenario and in the 3x3 blocks in the center of the urban scenario.

V2X communications are simulated using the network simulator ns-3 that is widely used in V2X communications research. All vehicles communicate using the ITS-G5 V2X standard (based on IEEE 802.11p) and therefore transmit using the CSMA/CA (Carrier Sense Multiple Access with Collision Avoidance) protocol. The propagation effects are modeled using the Winner+ B1 propagation model. Winner+ B1 differentiates between Line-of-sight (LOS) and Non-line-of-sight (NLOS) propagation conditions, and hence allows us to consider the strong impact of buildings in urban scenarios on the V2X communications performance. Following [13], the Winner+ B1 model has been adapted for V2V communications by configuring the antenna height to 1.5 m. The transmission power is set to 23 dBm and the packet sensing threshold to -85 dBm. All vehicles transmit using the 6 Mbps data rate (i.e. they utilize QPSK modulation with ½ code rate) and the channel bandwidth and carrier frequency are set to 10 MHz and 5.9 GHz, respectively. The ns-3 simulator has been extended with a CPS component and a sensing module implemented by the authors. The CPS component creates CPM messages based on ETSI's CPM message format [11]. CPM messages are generated following the ETSI CPM generation rules (Section II). The $T\_GenCpm$ is set to 0.1 s to enable the rapid transmission of newly detected objects and avoid long delays in the transmission of previously detected objects. Vehicles are equipped with a 360º sensor with a sensing range of 150 m [11].

We evaluated the performance of the current ETSI CPM generation rules in the urban and highway scenarios. The evaluation showed that the existing rules generate on average 9.8 and 9.6 CPMs per second per vehicle in the low and high traffic density highway scenarios, respectively. These results reveal that most CPMs are generated every 100 ms independently of the traffic density. In the urban scenario, the average number of CPMs generated per second per vehicle for the low and high traffic density scenarios is equal to 6.1 and 5.7, respectively. In the urban scenario, the CPM generation interval varies between 100 ms and 1 s. This is due to larger variations in the speed of vehicles and in the number of objects detected by each vehicle in the urban scenario (e.g. vehicles tend to concentrate at intersections) than in the highway scenario.

Figure 2 shows the PDF (Probability Density Function) of the number of detected objects by each vehicle and the number of objects included in each CPM when considering the ETSI CPM generation rules in the urban and highway scenarios. The obtained results show that the number of detected objects is non-negligible in both scenarios. The obtained results also show that around 50%-60% of the CPMs contained 4 or less objects in the highway scenario. The figure also reveals that around 50% of the CPMs generated in the urban scenario contained only 1 object while around 90% contained 3 or less objects. The obtained results demonstrate that the number of objects included in each CPM is significantly lower than the number of detected objects in both urban and highway environments. These results clearly confirm the problem previously described and illustrated in Figure 1 for realistic scenarios: the ETSI CPM generation rules generate frequent CPMs that contain a small number of detected objects.

The transmission of frequent and small CPMs adds significant overhead. This overhead increases the channel load and can reduce the reliability of V2X communications and thus degrade the perception of CAVs. To overcome these challenges, we propose an improved algorithm that avoids the frequent transmission of CPMs with a small number of objects.

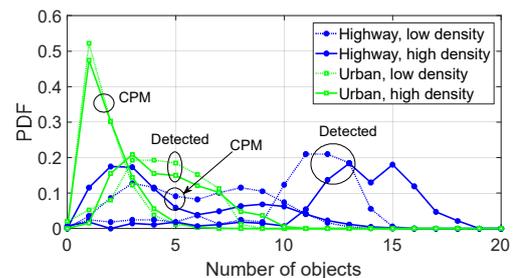

Figure 2. PDF of the number of objects detected by each vehicle and included in each CPM with the ETSI CPM generation rules.

## IV. Look-ahead proposal

Our *look-ahead* proposal is designed with the objective to reduce the channel load generated by CPMs while improving the perception capabilities of CAVs. To this aim, we propose a simple yet effective improvement of the current ETSI CPM generation rules to combat its challenges previously discussed. It was a design objective to minimize the changes to the ETSI proposal for higher standardization impact.

In our proposal, vehicles check the conditions to generate a new CPM every *T_GenCpm* following ETSI generation rules. Following these rules, we compute for each detected object the difference in absolute position ($\Delta P$), speed ($\Delta S$) and time elapsed ($\Delta T$) since the last time the detected object was included in a CPM. A new CPM is generated if at least one of the three conditions specified in Section II is satisfied. In other words, the CPM must include the information about the detected objects that satisfy $\Delta P$>4 m or $\Delta S$>0.5 m/s or $\Delta T$>1 s. These are the original conditions of the ETSI rules that we maintain in our proposal. This ensures that our proposal includes each object in a CPM at least as frequently as the ETSI rules. The pseudo-code for this process is shown in lines 1-8 of Algorithm I.

---

ALGORITHM I.
Input: Detected objects / Output: Objects (if any) to include in CPM
Execution: Every *T_GenCpm*

1. Set *flag* = false
2. **For** every detected object **do**
3.    Calculate $\Delta P$, $\Delta S$ and $\Delta T$ since the last time included in a CPM
4.    **If** $\Delta P$>4 m || $\Delta S$>0.5 m/s || $\Delta T$>1 s **then**
5.      Include object in current CPM
6.      Set *flag* = true
7.    **End If**
8. **End For**
9. **If** *flag* = true **then**
10.   **For** every detected object not included in current CPM **do**
11.    Calculate *Next* $\Delta P$, *Next* $\Delta S$ and *Next* $\Delta T$
12.    **If** *Next* $\Delta P$>4 m || *Next* $\Delta S$>0.5 m/s || *Next* $\Delta T$>1 s **then**
13.      Include object in current CPM
14.    **End if**
15.   **End For**
16. **End If**

---

Our proposal is triggered every time a new CPM must be generated by the ETSI rules. Then, our algorithm looks ahead and predicts if any of the detected objects that are not included in the current CPM would be included in the following CPM. The prediction is computed as follows considering that the objects maintain their current acceleration:

$$Next\ \Delta P = \Delta P + S \cdot T\_GenCpm + 0.5 \cdot A \cdot T\_GenCpm^2 \quad (1)$$

$$Next\ \Delta S = \Delta S + A \cdot T\_GenCpm \quad (2)$$

$$Next\ \Delta T = \Delta T + T\_GenCpm \quad (3)$$

where *S* and *A* are the current speed and acceleration of the detected object. Our algorithm includes in the current CPM (instead of the following one) the detected objects that satisfy *Next* $\Delta P$>4 m or *Next* $\Delta S$>0.5 m/s or *Next* $\Delta T$>1 s. This CPM includes the current information about these objects. Anticipating the inclusion of a detected object in a CPM is proposed to avoid transmitting many CPMs with information about a small number of detected objects. The proposed algorithm is robust against prediction errors resulting from the irregular movement of the detected objects since the worst-case prediction scenario will result in our proposal operating like the ETSI CPM generation rules. The pseudo-code for this anticipatory extension of the ETSI CPM generation rules is described in lines 9-16 of Algorithm I.

## V. Evaluation

This section compares our proposal with the ETSI CPM generation rules considering the same highway and urban simulation scenarios described in Section III. In the simulations, vehicles detect objects using their onboard sensors and CPMs are generated following the conditions in Section II.

### A. Generation of CPMs

We first analyze how our proposal influences the generation of CPMs. In particular, we study how it impacts the CPM generation rate and the number of objects contained in each CPM. Table I compares the average number of CPMs generated per second per vehicle and the number of objects (i.e. vehicles) per CPM with our proposal and with the ETSI CPM generation rules. The table also reports the difference between the two algorithms. Table I shows that our proposal reduces (between 33% and 44%) the number of CPMs generated per second compared to the ETSI rules. This reduction is achieved by anticipating the transmission of information about detected objects and increasing the number of objects included in each CPM. Table I shows that our proposal augments (between 63% and 110%) the average number of objects included in each CPM in urban and highway scenarios. The improvement is higher in the highway scenario because CPMs are often sparsely transmitted (around 30% above 300 ms) in the urban scenario and each vehicle detects less objects. These effects make it more difficult to group the information about detected objects in less CPMs in the urban scenario.

TABLE I. AVERAGE CPM RATE AND NUMBER OF OBJECTS IN EACH CPM

| Traffic Density | Algorithm | CPM rate | | Number of objects | |
|---|---|---|---|---|---|
| | | Highway | Urban | Highway | Urban |
| Low | ETSI | 9.8 Hz | 6.1 Hz | 6.1 | 1.7 |
| | Look-ahead | 6.0 Hz | 4.1 Hz | 11.9 | 2.9 |
| | Difference | -38.8 % | -32.8 % | +95.1 % | +70.6 % |
| High | ETSI | 9.6 Hz | 5.7 Hz | 5.1 | 1.9 |
| | Look-ahead | 5.4 Hz | 3.9 Hz | 10.7 | 3.1 |
| | Difference | -43.8 % | -31.6 % | +109.8 % | +63.2 % |

### B. V2X communications performance

The previous section has shown that our proposal reduces the number of CPMs by augmenting the number of objects reported in each CPM. This reduces the communications overhead and decreases the channel load that is here measured with the CBR (Channel Busy Ratio). The CBR is the percentage of time that the channel is sensed as busy and is calculated as in [14]:

$$CBR = T_{busy}/T_{CBR} \quad (4)$$

where $T_{busy}$ is the time (in milliseconds) during which the strength of received signals exceeds -85 dBm. $T_{busy}$ is computed over a period of $T_{CBR}$ = 100 ms. Table II shows that our proposal reduces the CBR between 10% and 23% depending on the scenario and traffic density. This reduction results from transmitting less CPMs and consequently reducing the communications overhead. The reduction of CBR is higher in the urban scenario because CPMs include information about a lower number of objects than in the highway scenario. As a result, the communications overhead represents a larger portion of the transmitted bits in the urban scenario (76% with ETSI rules) than in the highway scenario (49%). These results demonstrate that our proposal reduces the channel load in both scenarios and hence improves the system's scalability.

TABLE II. AVERAGE CBR AND MAXIMUM DISTANCE WITH PDR ≥ 0.9

| Traffic Density | Algorithm | CBR | | Distance | |
|---|---|---|---|---|---|
| | | Highway | Urban | Highway | Urban (LOS) |
| Low | ETSI | 29.2 % | 12.7 % | 132 m | 182 m |
| | Look-ahead | 26.1 % | 9.7 % | 151 m | 205 m |
| | Difference | -10.6 % | -23.6 % | +14.4 % | +12.6 % |
| High | ETSI | 49.4 % | 19.9 % | 102 m | 134 m |
| | Look-ahead | 41.4 % | 15.6 % | 118 m | 162 m |
| | Difference | -16.2 % | -21.6 % | +15.7 % | +20.9 % |

Reducing the CBR and channel load reduces the packet collisions and improves the reliability of V2X communications. We measure this reliability using the PDR (Packet Delivery Ratio) metric that is defined as the probability of correctly receiving a packet at a given distance $d$ to the transmitter. The PDR is calculated for a given transmitting vehicle $j$ as:

$$PDR_j(d) = \frac{\sum_{i=1}^{N} X_{i,j}(d)}{\sum_{i=1}^{N} Y_{i,j}(d)} \quad (5)$$

where $Y_{i,j}(d)$ is the number of vehicles that are located at a distance between $d-\Delta D/2$ and $d+\Delta D/2$ to the transmitter when the transmitter transmits packet $i$. $X_{i,j}(d)$ is the number of vehicles that successfully receive such packet $i$. $N$ denotes the number of transmitted messages and $\Delta D$=25 m. Each value of PDR($d$) corresponds to the average PDR at $d$ for all transmitting vehicles $j$. Figure 3 depicts the PDR achieved with the ETSI CPM generation rules and our proposal in the urban scenario under low and high traffic densities. The figure plots the PDR under LOS and NLOS propagation conditions between transmitter and receiver. Figure 3 shows that our proposal improves the reliability of V2X communications under LOS thanks to the reduction of the channel load and packet collisions[2]. Under NLOS conditions, the PDR is significantly degraded as it is mostly affected by the propagation conditions due to the presence of buildings. In NLOS conditions, reducing the communications overhead with our proposal does not have a significant positive impact on the PDR.

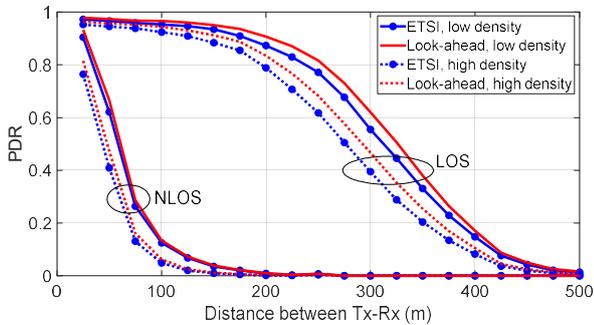

Figure 3. PDR (Packet Delivery Ratio) for the urban scenario.

---

[2] Similar trends are observed in the highway scenario.

The PDR has a direct impact on the V2X communications range. Table II reports the distance up to which a PDR equal or higher than 0.9 is guaranteed. 3GPP considers this distance as a reference V2X performance metric [13]. Table II shows that our proposal increases this distance compared to the current ETSI CPM generation rules by 12% to 20%. These results show that our proposal increases the reliability of V2X communications thanks to the reduction of the channel load.

### C. Perception capabilities

The previous sections have shown that our proposal improves the V2X communications performance due to the reduction of the channel load resulting from the reorganization of CPMs. This section evaluates how our proposal impacts the perception capabilities of CAVs. We measure the perception capabilities using the Object Perception Ratio (OPR) metric that is defined as the probability to detect an object within a given time window $\Delta T$ thanks to the exchange of CPMs. We consider that a vehicle successfully detects an object if it receives at least one CPM with information about that object during $\Delta T$. The time window has been set equal to the time required by the CPM generation rules for a vehicle to send an update about a detected object considering the speed of the object. The time window $\Delta T$ is dynamically computed for each object based on its speed $S$ as $\Delta T = T\_GenCpm \cdot [4 \cdot S^{-1} \cdot T\_GenCpm^{-1}]$, with $\Delta T \leq 1$ s. This computation considers that an object moving at speed $S$ is included in a CPM every time it has moved 4 m, and that the CPM period is a multiple of $T\_GenCpm$. Considering this dynamic adaptation of the time window, the OPR metric of vehicle $i$ and object $j$ is:

$$OPR_{i,j}(d) = \frac{S_{i,j}(d)}{T_{i,j}(d)} \quad (6)$$

$T_{i,j}(d)$ is the time during which object $j$ is located at a distance between $d-\Delta D/2$ and $d+\Delta D/2$ from vehicle $i$. $S_{i,j}(d)$ is the time during which vehicle $i$ has successfully detected object $j$ and their distance was between $d-\Delta D/2$ and $d+\Delta D/2$. $S_{i,j}(d)$ is computed taking into account the CPMs received during $T_{i,j}(d)$. Note that $S_{i,j}(d) \leq T_{i,j}(d)$. The OPR metric at a distance $d$ is computed as the average value of $OPR_{i,j}(d)$ for all vehicles $i$ and all objects $j$. $\Delta D$ has been set equal to 25 m.

Figure 4 plots the OPR metric as a function of the distance between the detected object and the vehicle receiving the CPMs. In the urban scenario, we differentiate the cases where the detected object and the vehicle receiving the CPMs are in the same street or in a perpendicular street. This helps us estimate the effectiveness of collective perception as a function of the relative position of the detected object to the vehicle receiving the CPMs, including whether they are under LOS or NLOS conditions. Figure 4 shows that our proposal improves the object perception ratio compared to the ETSI CPM generation rules in both highway and urban scenarios. This is due to two main reasons: 1) our proposal increases the PDR and therefore the probability to correctly receive CPMs, 2) our proposal reorganizes the transmission of detected objects in CPMs. This reorganization increases the average number of times that a detected object is reported in a CPM compared to the ETSI generation rules (by 20% and 10% in the highway and urban scenarios, respectively). This increases the probability to receive information about a detected object and hence the OPR.

Figure 4 shows that the highest perception levels are achieved in the highway scenario where our proposal also obtains its highest improvement compared to the ETSI generation rules. In the urban scenario, buildings significantly attenuate the radio signal and block the sensors field of view. High perception levels can hence only be achieved when the object and the vehicle receiving the CPMs are in the same street. However, the object perception ratio under these conditions is still lower in the urban scenario than in the highway one. This is the case because the urban scenario has lower traffic densities, and consequently, less vehicles detect and report information about each object. Figure 4 also shows that the object perception ratio is significantly degraded (independently of the generation rules) in the urban scenario when the object and the vehicle receiving CPMs are in perpendicular streets. This is because the object and the transmitting vehicle must be in the same street, and thus the transmitting and receiving vehicles are under NLOS conditions (unless the transmitting vehicle is at an intersection). These conditions significantly degrade the PDR and reduce the probability to receive CPMs.

We also analyze the perception capabilities of CAVs by computing the average time between updates that a vehicle receives about a detected object. The updates can be received from any vehicle that has detected the same object. A lower time between updates improves the perception since a vehicle receives more frequently information about a detected object. Figure 5 plots the average time between updates as a function of the distance between the object and the vehicle receiving the CPMs. Figure 5 shows that our proposal reduces the time between updates compared to the ETSI rules, especially at high distances. This is important because the perception capabilities of onboard sensors decrease with the distance. This improvement is achieved in both highway and urban scenarios independently of the traffic density.

VI. CONCLUSIONS

Cooperative or collective perception improves the perception capabilities of connected and automated vehicles. ETSI has proposed to date the first set of message generation rules for collective perception. These rules have a strong impact on perception since they define when collective perception messages should be generated and transmitted. This study shows that the current message generation rules for collective perception create frequent collective perception messages, and each message reports only about a few detected objects. This increases the communications overhead and degrades the V2X reliability as well as the perception capabilities. This paper proposes an improved algorithm for the message generation rules in collective perception. The proposal reduces the number of collective perception messages per second by reorganizing how information about detected objects is transmitted. Our proposal is able to simultaneously reduce the communications load and overhead, and improve the reliability of V2X communications and the perception of CAVs. This is achieved by reorganizing the transmission and content of CPMs.

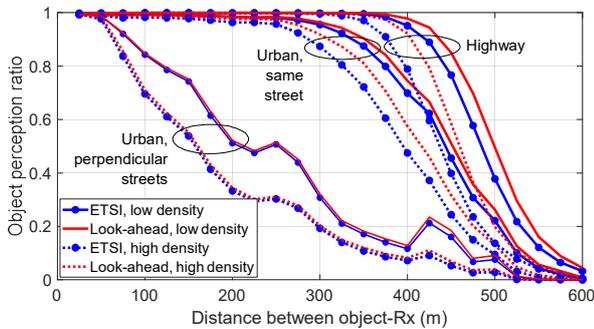

Figure 4. Object perception ratio as a function of the distance between the detected object and the vehicle receiving the CPM.

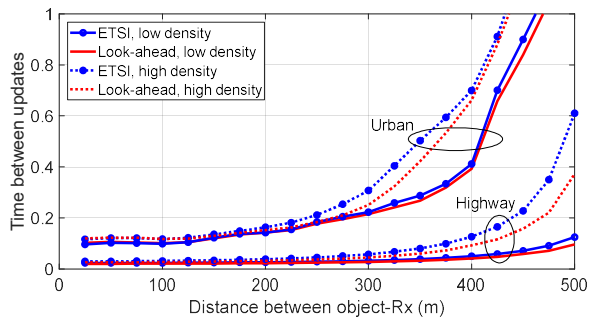

Figure 5. Average time between updates as a function of the average distance between the detected object and the vehicle receiving the CPMs.